# Application of silicene, germanene and stanene for Na or Li ion storage: A theoretical investigation

Bohayra Mortazavi[*,1], Arezoo Dianat[2], Gianaurelio Cuniberti[2], Timon Rabczuk[1,#]

[1]*Institute of Structural Mechanics, Bauhaus-Universität Weimar, Marienstr. 15, D-99423 Weimar, Germany.*

[2]*Institute for Materials Science and Max Bergman Center of Biomaterials, TU Dresden, 01062 Dresden, Germany*

## Abstract

Silicene, germanene and stanene likely to graphene are atomic thick material with interesting properties. We employed first-principles density functional theory (DFT) calculations to investigate and compare the interaction of Na or Li ions on these films. We first identified the most stable binding sites and their corresponding binding energies for a single Na or Li adatom on the considered membranes. Then we gradually increased the ions concentration until the full saturation of the surfaces is achieved. Our Bader charge analysis confirmed complete charge transfer between Li or Na ions with the studied 2D sheets. We then utilized nudged elastic band method to analyze and compare the energy barriers for Li or Na ions diffusions along the surface and through the films thicknesses. Our investigation findings can be useful for the potential application of silicene, germanene and stanene for Na or Li ion batteries.

*Keywords: Silicene; germanene; stanene; first-principles; Li ions;*

*Corresponding author (Bohayra Mortazavi): bohayra.mortazavi@gmail.com

Tel: +49 157 8037 8770,

Fax: +49 364 358 4511

#Timon.rabczuk@uni-weimar.de

## 1. Introduction

The interest toward two-dimensional (2D) materials was raised by the great success of graphene [1–3]. Graphene is a zero-gap semiconductor that present outstanding mechanical [4] and heat conduction [5] properties, surpassing all known materials. Graphene wide application prospects motivated experimental researches for the of other two-dimensional (2D) compounds, such as hexagonal boron-nitride [6,7], silicene [8,9], germanene [10], stanene [11], transition metal dichalcogenides [12–14] and phosphorene [15,16]. One of the most attractive areas for 2D materials research lies to their potential for integration. In response to high demand for more efficient rechargeable energy storage systems, graphene and other 2D crystals and their hybrid structures can be considered as one of the promising approaches [17–20]. Silicon based anodes due to the high specific capacity of 4200 mAh/g [21,22] are considered as an alternative to the conventional graphite based anode materials. The degradation of silicon and its drastic volume changes during the battery operation nevertheless question the commercialization of this battery technology [23–25]. Having similar atomic and electronic structures as that of the silicon, germanium and tin are also expected to present high capacity for Li-ion batteries. Silicene, germanene and stanene, the 2D form of these element, owing to their large surface area, high mechanical flexibility and higher electron mobility can be also therefore considered as interesting candidates to use in secondary batteries [20]. Nonetheless, because of experimental complexities, theoretical studies can be considered as an alternative to evaluate the application prospect [25–29]. This study consequently aims to probe the applicability of silicene, germanene and stanene for Li or Na ions storage using DFT calculations.

## 2. Modelling

All DFT calculations in the present study were carried out using Vienna ab initio simulation package (VASP) [30,31] using the Perdew-Burke-Ernzerhof (PBE) generalized gradient approximation exchange-correlation functional [32]. The projector augmented wave method [33] was employed with an energy cutoff of 450 eV. In this work, van der Waals interactions have been included using the semiempirical correction of Grimme [34], as it is implemented in VASP. This method have been shown to be a valid choice for calculating the binding energies [35,36]. We applied periodic boundary conditions in all directions with a 17 Å vacuum layer to avoid image-image interaction along the sheet thickness. In order to simulate adatoms adsorption, we included 60 atoms for the pristine silicene, germanene and stanene sheets. The planar size of these constructed periodic cells were about 20 Å×19.24 Å, 20.85 Å×20.07 Å and 23.92 Å×23 Å, for silicene, germanene and stanene membranes, respectively. We first identified the strongest binding sites for Na or Li ions and then we gradually increased the ions concentrations by randomly but uniformly locating them on the pre-identified sites. After automatically positioning the ions, conjugate gradient method energy minimization was performed with a $10^{-4}$ eV criteria for energy convergence using a 5×5×1 k-point mesh size for Brillouin zone sampling. A single point calculation was finally achieved to report the energy and electronic density of states (DOS) of the system at which the Brillouin zone was sampled using a 13×13×1 k-point mesh size. Tetrahedron method was employed for the calculation of DOS. For the evaluation of charge transfer between the ions and substrate, we performed Bader charge analysis [37]. Climbing-image nudged elastic band (NEB) [38] method was utilized to obtain diffusion pathways and the corresponding energy barriers. For the NEB calculations, we included 12 atoms for the films with a Monkhorst Pack mesh [39] of 15×15×1 special points. We note that we also compared the energy of the several systems with different number of k-points along the sheets thickness and we found that a single k-point yields absolutely accurate results.

Silicene [8,9], germanene [10] and stanene [11] are atomic thick materials with hexagonal honeycomb structures similar to that of graphene. Contrary to graphene, silicene, germanene and stanene are not flat, but have a periodically buckled topology. In Fig. 1, a periodic structure of silicene that was used to evluate Li or Na adatoms diffusion, with displayed buckled structure is illustrated. To define atomic structure of silicene, germanene and stanene, one only needs to define the lattice constant ($\alpha$) and the buckling high ($\Delta$). After the energy minimization and geometry optimization, the obtained lattice parameters are shown in table 1. As it can be observed, we found good agreements with available information in the literature.

Table 1, Lattice constant ($\alpha$) and the buckling high ($\Delta$) of silicene, germanene and stanene.

|  | Lattice constant, $\alpha$ (Å) | | Buckling high, $\Delta$ (Å) | |
| --- | --- | --- | --- | --- |
|  | Present study | Previous reports | Present study | Previous reports |
| Silicene | 3.85 | 3.86[40], 3.83[41] | 0.46 | 0.45[40], 0.44[9], 0.44[41] |
| Germanene | 4.01 | 4.045[42], 3.97[41] | 0.7 | 0.68[42], 0.64[41] |
| Stanene | 4.67 | 4.67[11], 4.68[42] | 0.86 | 0.85[11], 0.86[42] |

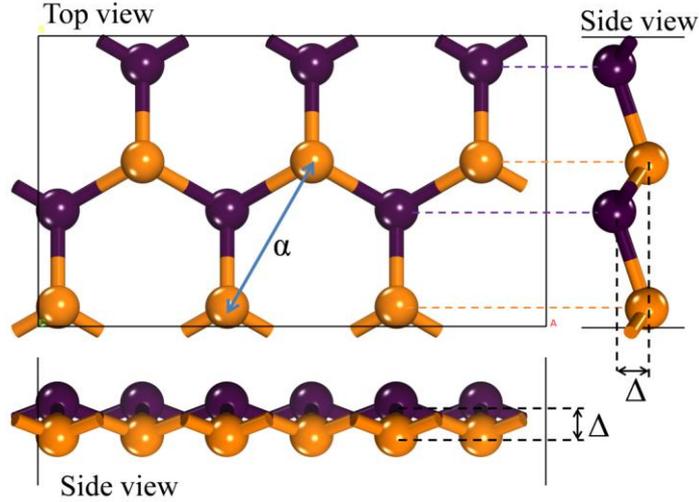

Fig-1, Atomic structure of silicene, germanene or stanene can be defined by the hexagonal lattice constant (α) and the buckling high (Δ). Silicene, germanene or stanene are atomic thick materials that are only made from Si, Ge or Sn atoms, respectively. Here, in order to facilitate the visualization of these buckled sheets, atoms with the same out of plane atomic positions are colored in either orange and purple.

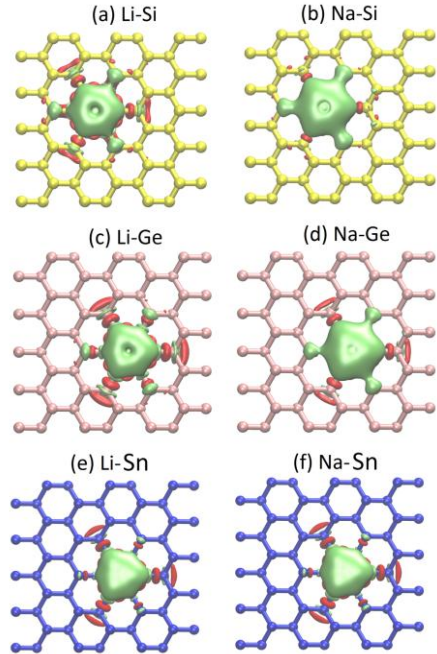

Fig-2, Calculated charge density difference plots for Li or Na ions on silicene, germanene and stanene. The green and red areas represent electron losses and gains, respectively.

## 3. Results and discussions

We first evaluate the most stable adsorption sites for a single Na or Li atoms on silicene, germanene and stanene sheets. A relatively large energy for the ion adsorption on an anode material is a highly demanding factor. To find the most stable adsorption site, we considered three different initial positions: above the top of an atom, above the middle-point of a bond and finally above the hollow site in the hexagonal lattice. In order to accurately find the adsorption site, we used a relatively strict energy criteria of $10^{-5}$ eV for the conjugate gradient method energy minimization termination. Interestingly, for the both Na and Li atoms on silicene, germanene and stanene, we found that the hollow sites present the maximum binding energy. This observation is in agreement with previous studies for Li atom adsorption on silicene [20,43]. For these hollow sites, in Fig. 2 the calculated charge density difference plots for Li or Na ions are illustrated. It can be seen that electrons tend to accumulate around Si, Ge or Sn atoms and their density surrounding Na or Li atom decreases.

We then investigate the binding energy, $E_b$, which can be calculated using the following relation:

$$E_b = \frac{(E_{sub} + N \times E_{ion} - E_{sub+ion})}{N} \qquad (1)$$

where $E_{sub+ion}$ denotes the total energy of the structure, $E_{sub}$ is the total energy of the substrate layer, $E_{ion}$ is the energy of an isolated neutral adatom in vacuum and N is the number of adatoms. For the low concentration of Li atoms on silicene, we found a binding energy of around 2.1 eV. This value is remarkably close to the previous theoretical results of 2.08 eV [20], 2.21 [43] and 2.13 eV [44]. For Na adsorption on silicene, we found a binding energy of 1.58 eV which is also close to the predicted value of 1.61 eV by Lin and Ni [44]. Interestingly, binding energies of Li on germanene and stanene are also both predicted to be around 2.07 eV. Similarly, for Na atom adsorption on germanene and stanene, we predicted a binding energy of 1.54 eV. In Fig. 3, we plot the calculate binding energies as a function of adatoms concentrations. For the all studied cases, the binding energy remains almost constant with increasing the ions concentration. Theoretical capacity of an anode or cathode electrode material correlates to its molecular weight and also the efficiency of the electron transfer. Theoretical capacity can be calculated on the basis of the Faraday constant. We note that we performed the Bader charges analysis over the all samples and it was confirmed that all adsorbed atoms transferred successfully their single valance electron to the substrate which is equivalent to the charge capacity of 954 mAh/g, 369 mAh/g and 226 mAh/g for Li or Na ions storage using single-layer silicene, germanene and stanene sheets, respectively.

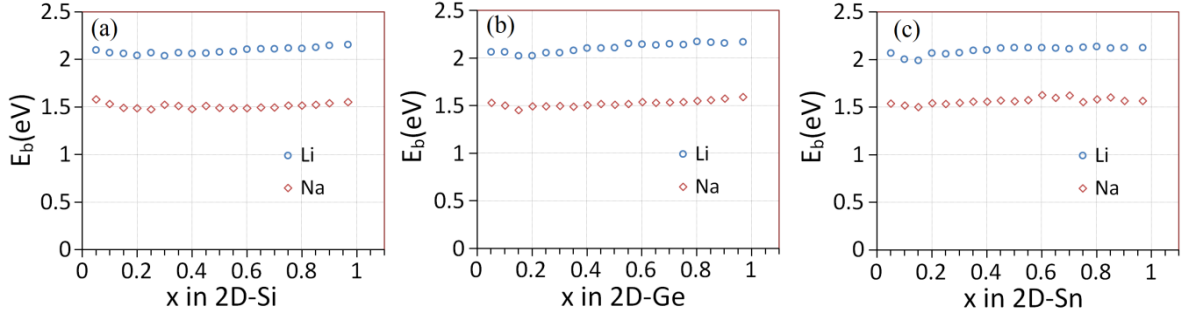

Fig-3, Concentration dependent binding energy for Li or Na ions on silicene, germanene and stanene. For the fully covered structures, x is equal to 1.

During frequent charging or discharging processes of a rechargeable battery, the ions concentration in solid particles change depending on the current direction. This process accordingly induce compositional and structural change of the electrode materials. In Fig. 4 we plot samples of silicene and stanene films covered with various concentration of Li adatoms. Results illustrated in Fig. 4, reveal that by increasing the atoms content, the anode films undergo deformation and lattice distortion. For Na or Li atoms adsorption on silicene, however an almost fully saturated structure (Fig. 4d) presents a more uniform structure in comparison with a film with lower concentration of adatoms (Fig. 4c). This can be explained due to the fact that for a fully covered film, the structure is more symmetrical. On another hand, for the stanene sheet covered with Li atoms, we found that in general by increasing the ions content the distortion increases (Fig. 4f and Fig. 4h). In this case as it is shown in Fig. 4e, due to a much larger hollow site, the Li atoms are adsorbed almost inside the film. By increasing the ions concentration, consequently the new Li ions push more the surrounding Sn atoms inducing higher distortions in the lattice. Therefore, for a particular adatom, if the strongest adsorption site is placing upper than the film, the lattice distortion upon the ion insertion reduces. We should remember that the amount of distortion also correlate to the stiffness of a material. For silicene, germanene and stanene, in general, the stiffness has been reported to reduce considerably with increasing atomic weights (from Si to Sn) [45,46]. Lower stiffness rigidity can also well explain the higher distortion of the lattices. Nevertheless, by increasing the adatoms concentration on the considered films, the buckling length increases gradually. We mention that Na atom because of a larger radius adsorbs on the hollow site, upper than the Li atom. For Na atoms consequently the change in the buckling length is lower. Because of the free surfaces and the flexibility of the considered films, the maximum stresses induced in the structures are always below 0.72 GPa.nm. This stress is clearly low, implying that the structures cannot be degraded mechanically upon the ions intercalation.

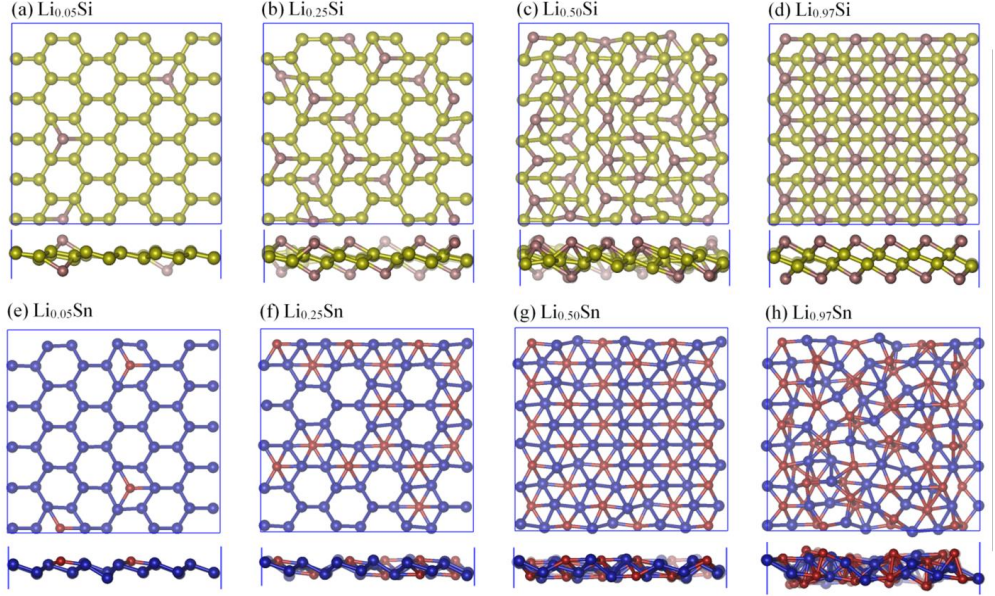

Fig-4, Samples of lithiated silicene and stanene films.

Diffusion of ions inside an anode plays an important role in the batteries performance. We next shift our attention to the diffusion pathway of a single Na or Li ions on silicene, germanene and stanene films. This was achieved by calculating the variations in energy as an adatom moves between equivalent adsorption sites. Diffusion process involves simultaneous motion of several atoms, including the Na or Li adatom and their Si, Ge or Sn neighbours [20]. Two representative diffusion pathways were considered, planar path which happens along the surface of the films and the other pathways is occurring along the sheet thickness by passing through the hexagonal hollow. In the all studied cases for the diffusion along the surface, first Li or Na atom at the hollow site jump over a Si, Ge or Sn atom, then they move to the next hollow site (Fig. 5a and Fig. 5c). Therefore, diffusion along the surface follows a zigzag pathway. The diffusion energy barrier for Na ion along the surface is predicted to be slightly smaller than that for Li ion. Na atom because of the larger radius is placed higher out of the hollow site and it can diffuse and facing less repulsive forces. For the both Na and Li ions, the diffusion is faster on the silicene surface and we found on the germanene and stanene the energy barrier is almost indifferent. On another side, for the through plane diffusion, silicene and stanene present the highest and lowest energy barriers, respectively. This can be explained because of a larger hexagonal hollow site for stanene. For example, for a Li atom diffusion through the plane, the diffusion barrier is 1.2 eV and 0.1 eV for silicene and stanene, respectively. For Na atoms, these barriers are much stronger, 3.56 eV, 2.19 eV and 0.8 eV on silicene, germanene and stanene, respectively. This can be attributed because of the larger atomic size of Na atoms that makes it harder to pass through the hexagonal hollow.

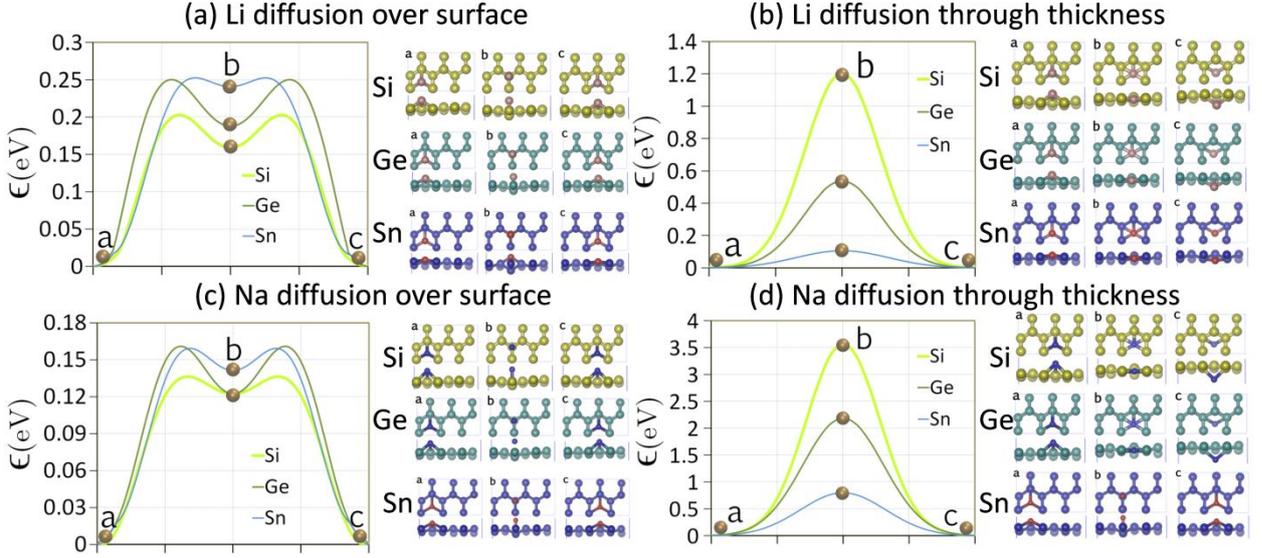

Fig-5, Comparison of diffusion energy barriers and the pathways for the Li or Na adatoms in silicene, germanene and stanene.

Electronic conductivity is among the important factors for the application prospect of an electrode material. The internal electronic resistance of a battery correlates strongly to the electronic conductivity of the electrode materials. Moreover, the ohmic heating generated during the charging or discharging process is also proportional to the electronic conductivity of the active material particles. We therefore briefly explore the electronic responses of studied systems. The calculated electronic DOSs are plotted in Fig. 6. For the all studied pristine films without Na or Li adatoms, at the zero state energy (Fermi level) the DOS is zero, this reveals zero-gap semiconducting electronic nature. Nevertheless, only in the case of silicene we found that upon lithiation, the DOS undergoes a transformation from a zero-gap to a semiconductor with a finite band gap. This finding is in agreement with the previous theoretical studies for lithiated silicene [20,43]. For the rest of studied systems, we found that upon the adatom adsorptions, total DOS for valence/conduction band around the Fermi level are increased. In these cases, at the Fermi level the band gap is disappeared, which implies that the structures are presenting metallic electronic response. It can be therefore concluded that only the silicene present semiconducting properties upon the Li atoms adsorption, which is not a desirable behaviour for its practical application as an anode material for Li ion batteries.

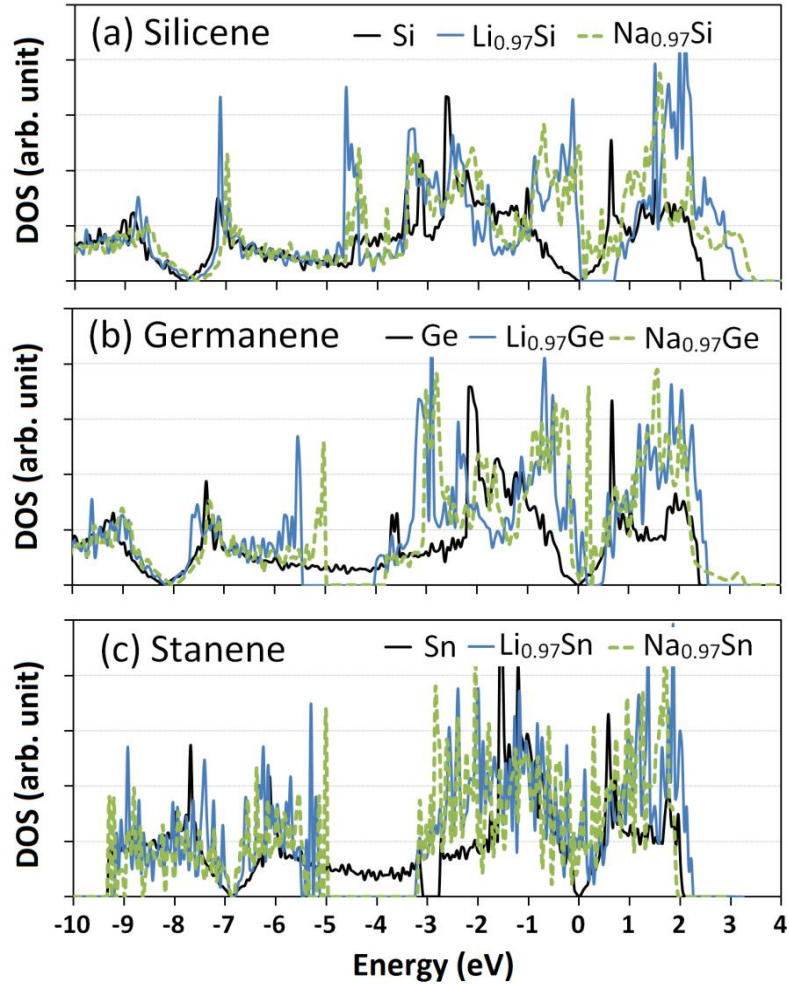

Fig-6, Electronic density of states (DOS) of silicene, germanene and stanene

### 4. Summary

In summary, we performed DFT calculations to investigate the application of silicene, germanene and stanene for Na or Li ion batteries. We found that in all cases the hexagonal hollow sites offer the maximum binding energies for ions. We predicted that the binding energy almost remains constant by increasing the ions concentration up to the full coverage of the films. Based on the Bader charge analysis results we predicted high charge capacities of around 954 mAh/g, 369 mAh/g and 226 mAh/g for Li or Na ions storage using single-layer silicene, germanene and stanene, respectively. We found that adatoms adsorption induce lattice distortion and in addition by increasing the ions coverage the buckling length of the sheets increases gradually. We estimated that the studied films structures cannot degrade mechanically upon the ions intercalation. We then probed the diffusion pathways and the corresponding energy barriers of Li and Na ions. We studied the diffusion pathways over the surface and also through the thickness of the films. Finally, we explored the electronic DOS of the studied films upon Li or Na ions adsorption. Our findings can be useful for the possible application of silicene, germanene and stanene for Na or Li-ion batteries.

## Acknowledgment

BM and TR greatly acknowledge the financial support by European Research Council for COMBAT project (Grant number 615132).